\documentclass[preprint2]{aastex}

\slugcomment{Submitted to the Astrophysical Journal}
\shorttitle{Shocked Molecular Gas Adjacent to W44}
\shortauthors{Sashida et al.}

\def\kms{\hbox{km s$^{-1}$}}
\def\VLSR{\hbox{$V_{\rm LSR}$}}

\def\dotdeg{\hbox{$.\!\!^\circ$}}

\begin{document}

\title{Kinematics of Shocked Molecular Gas Adjacent to the Supernova Remnant W44}
\author{Tomoro Sashida, Tomoharu Oka, Kunihiko Tanaka, Kazuya Aono, \&\ Shinji Matsumura}
\affil{Department of Physics, Faculty of Science and Technology, Keio University, 3-14-1 Hiyoshi, Kohoku-ku, Yokohama, Kanagawa 223-8522, Japan. }
\author{Makoto Nagai, \&\ Masumichi Seta}
\affil{Institute of Physics, University of Tsukuba, 1-1-1 Tennoudai, Tsukuba, Ibaraki 305-8571, Japan.}

\begin{abstract}
We mapped molecular gas toward the supernova remnant W44 in the HCO$^+$ {\it J} = 1--0 line with the Nobeyama Radio Observatory 45 m telescope and in the CO {\it J} = 3--2 line with the Atacama Submillimeter Telescope Experiment 10 m telescope.  High-velocity emission wings were detected in both lines over the area where the radio shell of W44 overlaps the molecular cloud in the plane of the sky.  We found that the average velocity distributions of the wing emission can be fitted by a uniform expansion model.  The best-fit expansion velocities are $12.2\!\pm\!0.3$ \kms\ and  $13.2\!\pm\!0.2$ \kms\ in HCO$^+$ and CO, respectively.  The non-wing CO {\it J} = 3--2 component is also fitted by the same model with an expansion velocity of $4.7\!\pm\!0.1$ \kms .  This component might be dominated by a post shock higher-density region where the shock velocity had slowed down.  The kinetic energy of shocked molecular gas is estimated to be $(3.5\!\pm\!1.3) \times 10^{49}$ erg.  Adding this and the energy of the previously identified HI shell, we concluded that $(1.2\!\pm\!0.2)\times 10^{50}$ erg has been converted into gas kinetic energy from the initial baryonic energy of the W44 supernova.  We also found ultra-high-velocity CO {\it J} = 3--2 wing emission with a velocity width of $\sim\!100$ \kms\ at $(l, b)\!=\!(+34\dotdeg 73 , -0\dotdeg 47)$.  The origin of this extremely high-velocity wing is a mystery.  
\end{abstract}

\section{Introduction}
The blast wave of a supernova (SN) explosion generates strong shocks in the interstellar medium, accelerating, compressing, and heating it, and forming a shell making up a supernova remnant (SNR).  At the same time, it releases an enormous baryonic energy of $E_{\rm bar} \sim 10^{51}$ erg.  Until its lifetime ends, most of the baryonic energy would be converted to radiative energy, and thus a small percentage of it remains as the kinetic energy ($E_{\rm kin}$) of interstellar matter (Chevalier 1974).  Recent extragalactic studies revealed that the conversion efficiency, $\eta \equiv E_{\rm kin}/E_{\rm bar}$, is $\sim\!0.1$ in quiescent star-forming galaxies, whereas it increases to $>\!0.4$ in starburst-like environments (Strickland \&\ Heckman 2009).  

The radio source W44 is a type II SNR having a distorted radio shell (e.g., Kundu \&\ Velusamy 1972; Clark et al. 1975; Reich et al. 1984; Kassim et al. 1992) and a center-filled X-ray morphology (e.g., Gronenschild et al. 1978; Smith et al. 1985; Rho et al 1994; Harrus et al. 1997).  It is located 3 kpc away from the sun (e.g., Radhakrishnan et al. 1972; Caswell et al. 1975), and its estimated age is $(0.65\mbox{--}2.5)\times 10^4$ yr (e.g., Smith et al. 1985; Harrus et al. 1997).  A giant molecular cloud (GMC) is associated with this SNR (Seta et al. 1998; Seta et al. 2004).  The detection of OH 1720 MHz maser emission (Claussen et al. 1997; Frail \&\ Mitchell 1998; Claussen et al. 1999), the enhanced CO {\it J} = 2--1/{\it J} = 1--0 ratio (Seta et al. 1998), and the compact spots of high-velocity wing emission (Seta et al. 2004) are evidence of violent interaction between the molecular cloud and the W44 SNR.  Extensive CO imaging has detected spatially extended moderately broad emission (SEMBE) which might also be molecular gas disturbed by the SNR (Seta et al. 2004).  

In this paper, we present extensive HCO$^{+}$ {\it J} = 1--0 and CO {\it J} = 3--2 maps of the W44 GMC (\S 3.1).  The HCO$^{+}$ {\it J} = 1--0 line traces high-density [$n(\mbox{H}_2)\gtrsim 10^5$ cm$^{-3}$) gas, whereas CO {\it J} = 3--2 traces mainly warm moderate-density [$T_{\rm k}\geq 30$ K, $n(\mbox{H}_2)\gtrsim 10^{3.5}$ cm$^{-3}$] gas.  We report the detection of faint spatially extended wing emissions in the HCO$^{+}$ {\it J} = 1--0 and CO {\it J} = 3--2 lines (\S 3.2).  The kinematics of shocked gas traced by the spatially extended wing emission is also presented.  These data enable us to estimate the kinetic energy of the shocked molecular gas and consequently give an estimate of the total kinetic energy that has been supplied by the W44 SNR.  We also report the discovery of ultra-high-velocity CO {\it J} = 3--2 wing emission (\S 3.3).

\section{Observations}
\subsection{NRO 45 m}
We mapped the W44 molecular cloud in the HCO$^+$ {\it J} = 1--0 line (89.1885 GHz) with the Nobeyama Radio Observatory (NRO) 45 m radio telescope in two ways.  One is a mapping of a small area in the wide frequency band (512 MHz) mode.  The objectives of this wide-band mapping were to confirm the detection of spatially extended HCO$^+$ wings and to investigate the spatial and velocity distribution of shocked molecular gas.  The other is an extensive mapping in the narrow frequency band (32 MHz) mode. The objective here was to investigate the detailed kinematics of the shocked gas traced by HCO$^+$ wing emission.  

In both mappings, the data were obtained with the 25-element focal plane receiver BEARS (Sunada et al. 2000).  At 90 GHz, the telescope has an FWHM beam size of 18\arcsec\ and a main beam efficiency ($\eta_{\rm MB}$) of 44\%.  During these observations, the system noise temperatures ranged between 250 and 400 K in a double sideband.  The antenna temperature was calibrated by the standard chopper-wheel method.  Pointing errors were corrected every 2 h by observing the SiO maser source R-Aql at 43 GHz.  The pointing accuracy of the telescope was good to $\leq\! 3\arcsec$ in both azimuth and elevation.  To correct the gain variation among the 25 beams of BEARS, we scaled the antenna temperature for each channel with refer to that taken with the single-beam SIS receiver S100, which is equipped with a quasi-optical image rejection filter.  A clean reference position of $(l, b)\!=\!(+33\dotdeg 750, -1\dotdeg 509 )$ was chosen.  We scaled the antenna temperature by multiplying by $1/\eta_{\rm MB}$ to get the main-beam temperature, $T_{\mathrm{MB}}$.  

The wide-band HCO$^+$ mapping observations were conducted in May 2007.  The mapping area includes the group of OH 1720 MHz maser spots in the southwestern edge of the W44 radio shell (Claussen et al. 1997).  The observations were performed by position switching to the clean reference position.  We used 25 sets of 1024 channel auto-correlators (ACs) in the 512 MHz bandwidth mode which has a 500 kHz resolution.  At 90 GHz, these correspond to a 1700 km s$^{-1}$ velocity coverage and a 1.7 \kms\ velocity resolution, respectively.  The data were reduced using the NEWSTAR reduction package.  We subtracted the baselines of the spectra by fitting linear lines.  The data were smoothed with a Gaussian function and resampled onto a $17\arcsec\times 17\arcsec\times 1.0$ \kms\ regular grid.  The rms noise level of the final map is 0.07 K in $T_{\rm MB}$ ($1\sigma$).  

The extensive HCO$^+$ mapping was conducted from April to May 2010.  The mapping area completely covers the radio shell of W44.  The observations were performed in the on-the-fly (OTF) mapping mode.  We used ACs in the 32 MHz-bandwidth mode which has a 37.8 kHz resolution; these values correspond to a 107 km s$^{-1}$ velocity coverage and a 0.1 \kms\ velocity resolution, respectively.  The data were reduced using the NOSTAR reduction package.  They were smoothed with a Bessel--Gaussian function and resampled onto a $17\arcsec\times 17\arcsec\times 1$ \kms\ regular grid.  The rms noise level of the final map is 0.075 K in $T_{\rm MB}$ ($1\sigma$).  

\subsection{ASTE}
We also observed the W44 molecular cloud in the CO {\it J}=3--2 line (345.795 GHz) using the Atacama Submillimeter Telescope Experiment (ASTE) 10 m telescope.  The observations were conducted in June and November 2011.  The telescope has a beam efficiency of 0.6 and a FWHM beam size of 22\arcsec\ at 345 GHz.  The pointing of the telescope was checked and corrected every 2 h by observing R-Aql, and its accuracy was maintained within 2\arcsec (rms).  The observations were made with a waveguide-type sideband-separating SIS mixer receiver CAT345 in the OTF mapping mode.  The reference position was taken at $(l, b)\!=\!(+33\dotdeg 75, -1\dotdeg 509 )$.  During the observations, the system noise temperatures ranged between 200 and 600 K (single sideband).  The antenna temperature ($T_{\rm A}^{*}$) was obtained by the standard chopper-wheel technique.  

We used the digital AC spectrometer MAC (Sunada et al. 2000) in the wide-bandwidth mode, which covers an instantaneous bandwidth of 512 MHz with a 0.5 MHz resolution.  At 346 GHz, this bandwidth and resolution correspond to a 444 km s$^{-1}$ velocity coverage and 0.43 \kms\ velocity resolution, respectively.  The full extent of the W44 radio shell was mapped in the OTF mode.  The obtained data were reduced using the NOSTAR reduction package.  Linear baselines were subtracted from the spectra.  We scaled the antenna temperature by multiplying by $1.667\, (=1/0.6)$ to obtain the main beam temperature $T_{\mathrm{MB}}$.  The data were smoothed with a Bessel--Gaussian function and resampled onto a $8.5\arcsec\times 8.5\arcsec\times 1.0$ \kms\ regular grid.  The rms noise level of the final map is 0.15 K in $T_{\rm MB}$ ($1\sigma$).  

\begin{figure*}[htbp]
\epsscale{2.0}
\plotone{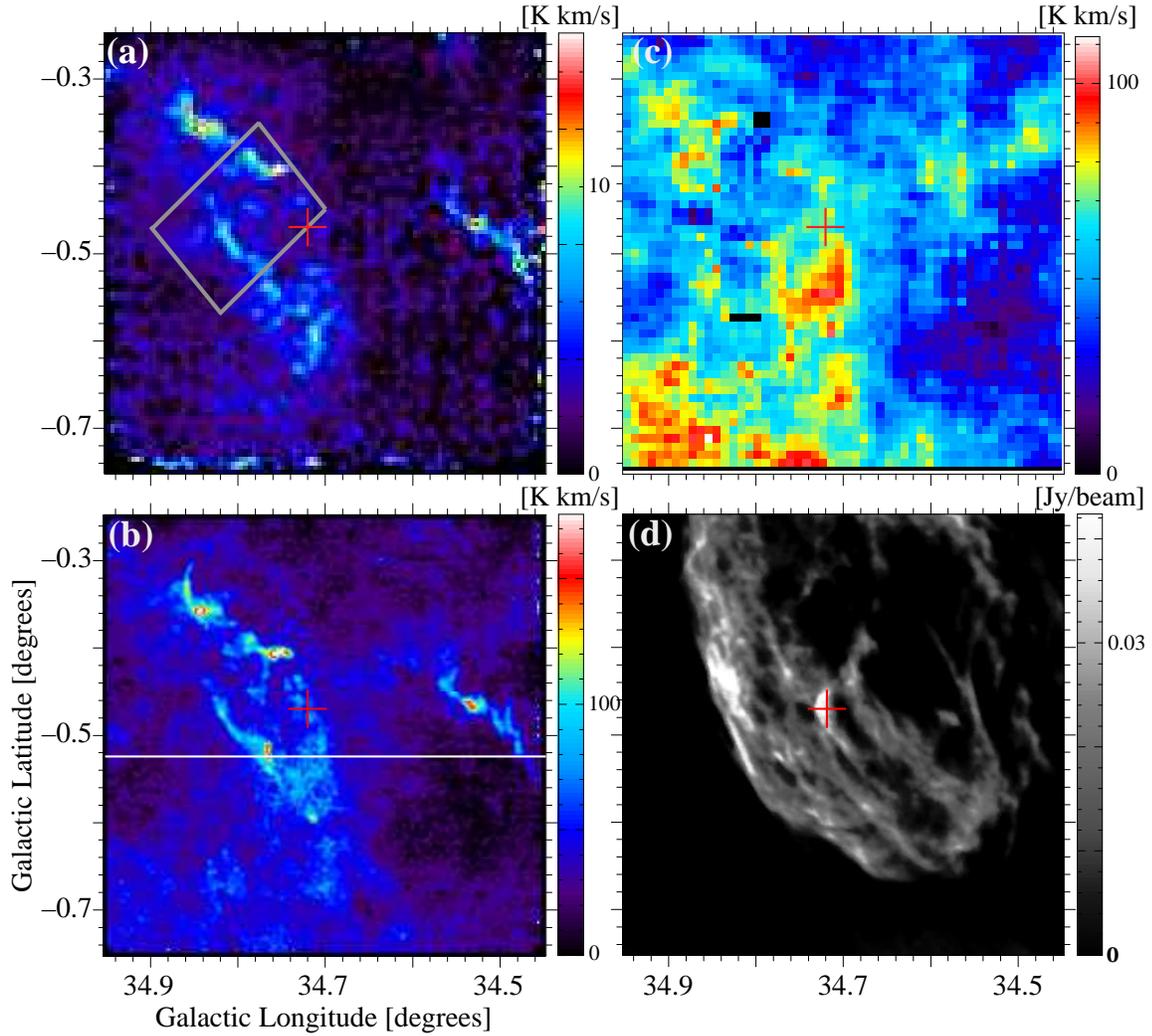}
\caption{({\it a}) Map of velocity-integrated HCO$^+$ {\it J} = 1--0 emission.  Rectangle indicates the area where wide-band observations were performed.   ({\it b}) Map of velocity-integrated CO {\it J} = 3--2 emission.  White solid line indicates $b=-0\dotdeg 524$, where longitude--velocity maps are shown in Fig.2.  ({\it c}) Map of velocity-integrated CO {\it J} = 1--0 emission.  The velocity range for integration extends from $\VLSR=+20$ \kms\ to $+60$ \kms\ for images ({\it a})---({\it c}).  ({\it d}) VLA radio continuum map at a wavelength of 20 cm (Jones et al. 1993).  Red cross in each panel indicates the position of the ultra-high-velocity wing emission (\S 3.3).
\label{fig1}}
\end{figure*}

\section{Results}
\subsection{Spatial Structure}
Figure 1 shows velocity-integrated maps of HCO$^+$ {\it J} = 1--0 and CO {\it J} = 3--2 emission.  These maps cover the entire extent of the SNR W44.  The VLA radio continuum image at 20 cm with a spatial resolution of $15\arcsec$ (Jones et al. 1993) is also placed on the right side for comparison.  In contrast to the widespread CO {\it J}=1--0 distribution (Seta et al. 2004), the HCO$^+$ {\it J} = 1--0 and CO {\it J} = 3--2 lines exhibit filamentary structure along with the radio continuum filaments of W44.  The HCO$^+$ {\it J} = 1--0 emission comes mainly from the area of the W44 radio shell, whereas the faint CO {\it J} = 3--2 emission is spread over the GMC in the Galactic-southeast of the shell.  At $(l, b)\!=\!(+34\dotdeg 57 , -0\dotdeg 43)$, another clump appears in both maps.  This clump has a velocity of $\VLSR \simeq +55$ \kms\ , which differs slightly from that of the associated GMC in the southeast.  This corresponds to the feature labeled C11 in Seta et al. (2004).  

\subsection{Velocity Structure}
Figure 2 shows longitude--velocity maps of the HCO$^+$ {\it J} = 1--0 and CO {\it J} = 3--2 emission.  The wide band and extensive mapping data are combined to draw the HCO$^+$ {\it l--V} map.  The broad velocity width feature at $(l, b)=(34\dotdeg 763, -0\dotdeg 524)$ is Wing 3 in Seta et al. (2004).  In addition to the previously detected wings, we discovered faint, spatially extended wing emission in both the HCO$^+$ {\it J} = 1--0 and CO {\it J} = 3--2 lines.  This extensive wing component completely overlaps the W44 radio shell and is therefore regarded as molecular gas shocked and disturbed by the SN blast wave.  This situation is well illustrated by Figure 3{\it a}, which shows the spatial distribution of the velocity dispersion calculated with the HCO$^+$ {\it J} = 1--0 data.  The velocity dispersion exceeds $10$ \kms\ only toward the radio shell, showing the spatial distribution of the extensive wing component.  

\begin{figure}[htbp]
\epsscale{0.8}
\plotone{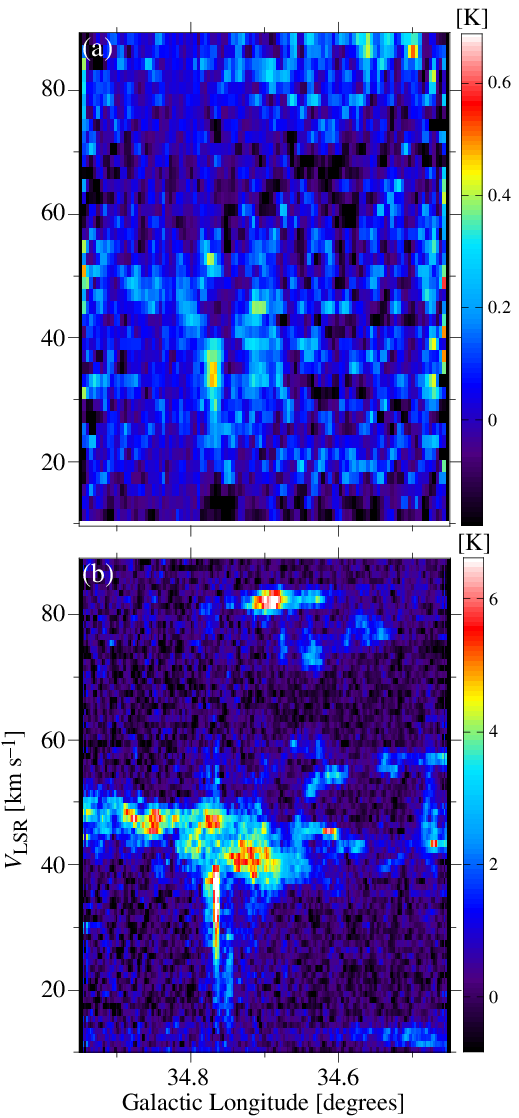}
\caption{({\it a}) Longitude--velocity map of HCO$^+$ {\it J} = 1--0 emission at $b=-0\dotdeg 524$.  Intense broad-line-width feature at $l=34\dotdeg 763$ is Wing 3 (Seta et al. 2004).  We also see faint broad-line-width feature at $(l, \VLSR )=(34\dotdeg 68\;\mbox{to}\;34\dotdeg 76, +15\,\kms \;\mbox{to}\; +55\,\kms)$.  ({\it b}) Longitude-velocity map of CO {\it J} = 3--2 emission.  The latitude is the same as for panel ({\it a}).  Wing 3 and a faint, spatially extended wing emission are apparent.  
\label{fig2}}
\end{figure}

We also show the spatial distribution of the average velocity calculated using the HCO$^+$ {\it J} = 1--0 data (Fig.3{\it b}).  The average velocity changes systematically from the edge to the center of the W44 radio shell.  The typical velocity at the southeastern edge is $\VLSR \simeq +45$ \kms, and that toward the center is $\simeq +35$ \kms; thus, the velocity difference is $\sim\!10$ \kms .  This velocity gradient from the edge to the center might be due to the expansion of the shocked gas.  In other words, shocked gas adjacent to the W44 SNR might be expanding at a velocity of $\sim 10$ \kms .  

\begin{figure}[htbp]
\epsscale{0.8}
\plotone{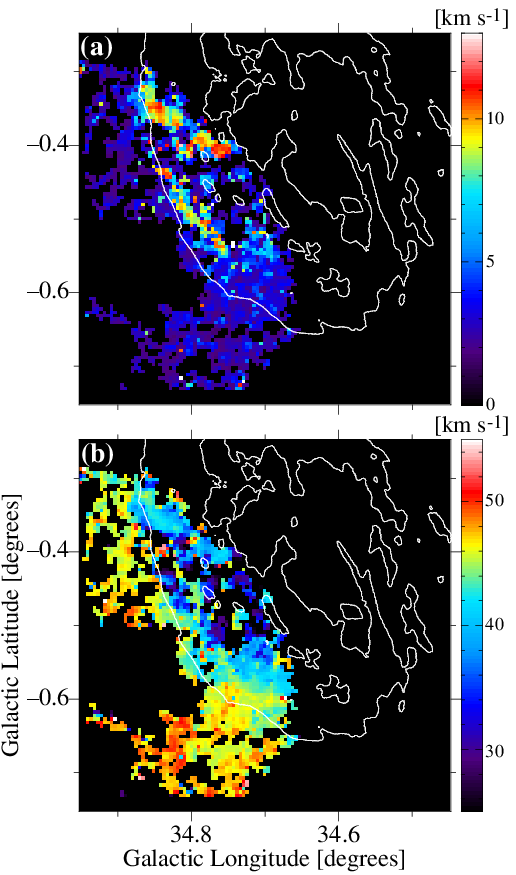}
\caption{({\it a}) Map of the velocity dispersion calculated with HCO$^+$ {\it J} = 1--0 data.  ({\it b}) Map of the average velocity calculated with HCO$^+$ {\it J}=1--0 data. Thick gray lines trace the outline of the W44 radio shell.  
\label{fig3}}
\end{figure}

\clearpage

\begin{figure*}[htbp]
\epsscale{2.0}
\plotone{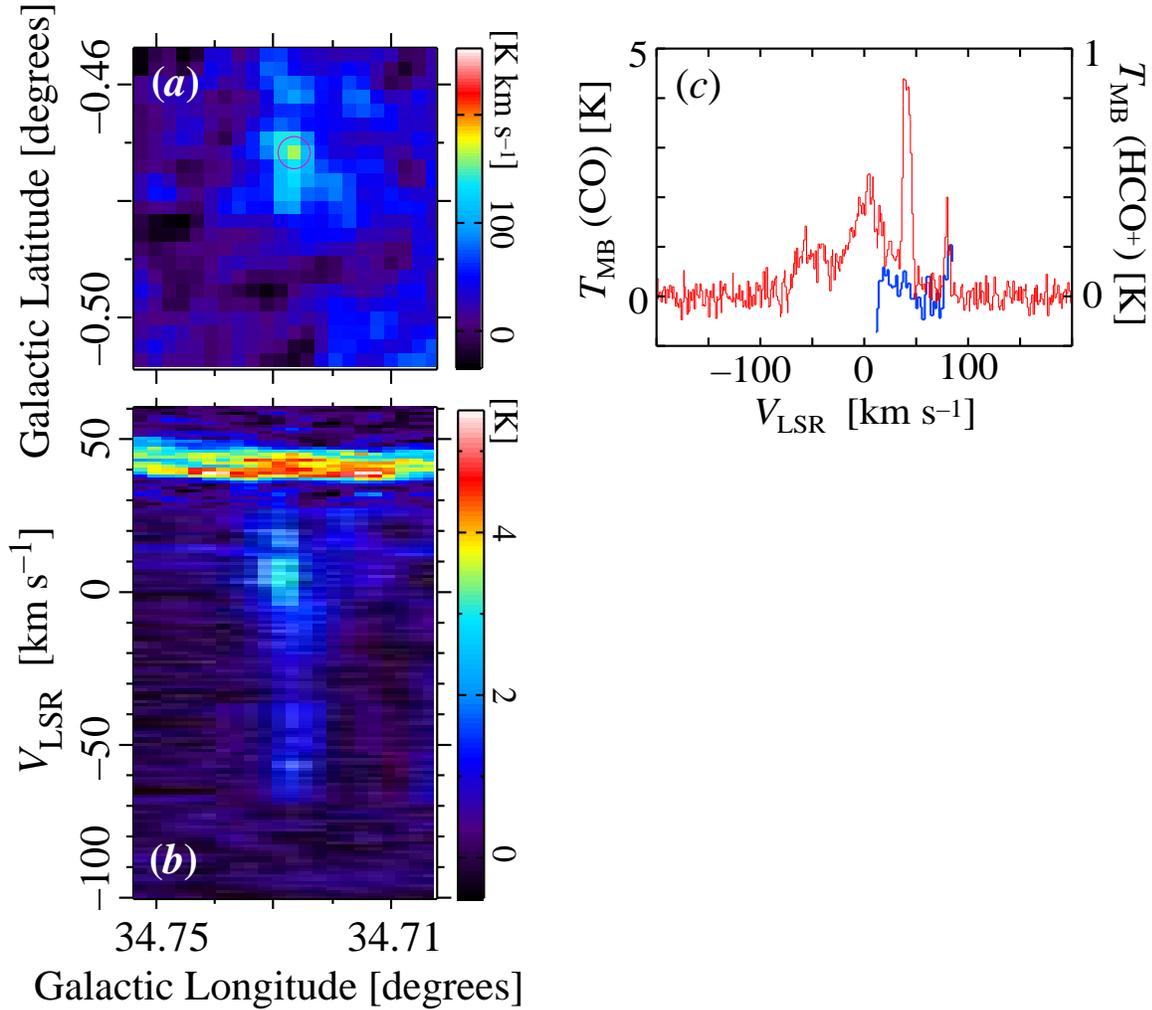}
\caption{({\it a}) Map of CO {\it J}=3--2 emission integrated over velocities from $-70$ \kms\ to $+30$ \kms.  Magenta circle indicates the area where the panel ({\it c}) spectra was created.  ({\it b}) Longitude-velocity map of CO {\it J}=3--2 emission at $b=-0\dotdeg 472$.    ({\it c}) Line profiles of CO {\it J}=3--2 (red) and HCO$^+$ {\it J} = 1--0 (blue) toward the ultra-high-velocity wing at $(l, b)\!=\!(+34\dotdeg 727 , -0\dotdeg 472)$.  Both spectra correspond to the emission smoothed spatially with a HPBW=20\arcsec\ Gaussian.  The HCO$^+$ profile is from the extensive mapping data.  
\label{fig4}}
\end{figure*}

\clearpage

\subsection{Ultra-high-velocity Wing}
At $(l, b)\!=\!(+34\dotdeg 73, -0\dotdeg 47)$, we noticed another high-velocity wing emission component of CO {\it J}=3--2 line with a velocity width of $\sim\!90$ \kms .  Figure 4{\it ab} shows the spatial and {\it l--V} distributions of this ultra-high-velocity wing emission.  The spatial size of this wing is very compact, $\sim\!0.5$ pc.  It arises from the main GMC component at $\VLSR \simeq +40$ \kms, as the other `normal' wings do, extending toward the negative velocity to $\VLSR = -70$ \kms.  Its velocity extension, $\Delta V\sim 100$ \kms, is far larger than that of the other wing emission detected in the W44 GMC.  Figure 4{\it c} shows lines profiles of CO {\it J}=3--2 and HCO$^+$ {\it J} = 1--0 toward the ultra-high-velocity wing.  The HCO$^+$ profile is made from the extensiive mapping data, since the wide-band mapping does not cover it unfortunately.  

Despite the great significance of the ultra-high-velocity-wing, its nature is somewhat controversial.  Spatially, it is intermingled with the extended wing emission.  A prominent blob of radio continuum emission seems to be associated with this wing (Fig.1{\it d}).  A nebulosity of H$_2$ 2.12 $\mu$m line emission also overlaps this wing (Fig.8 of Reach, Rho, \&\ Jarrett 2005).  These strongly suggest the close physical relationship between the ultra-high-velocity wing and the W44 SNR.  More research is necessary to reveal the origin of this peculiar wing emission.  

\section{Discussion}
\subsection{Velocity Components}
Detection of the extensive wing enables us to analyze the kinematics of shocked molecular gas adjacent to the W44 SNR.  As we saw in the previous section, its kinematics seems to be roughly reproduced by a simple expansion model.  Before analyzing the kinematics, we inspected the line profiles in each line to extract the high-velocity wing component correctly.  Hereafter, we refer to the high-velocity wing data as the ``wing" and to the data remaining after the wing is subtracted from the original data as ``non-wing data".  
The HCO$^+$ and CO data sets were analyzed separately.  

Figure 5 shows profiles of the HCO$^+$ {\it J}=1--0 and CO {\it J}=3--2 lines toward $(l, b)=(34\dotdeg 82, -0\dotdeg 47)$.  The HCO$^+$ {\it J}=1--0 line profiles suffer from absorption or additional emission at $\VLSR = +46$ \kms\ from the foreground quiescent gas, which belongs mainly to the associated GMC.  
We used the original HCO$^+$ data as the wing since the restoration of unabsorbed profiles was difficult.  In CO {\it J}=3--2 line profiles, the quiescent gas appears as narrow or moderate-velocity-width emission.  We referred to the CO {\it J}=1--0 data to subtract this quiescent gas component.  The subtracted main beam temperature is expressed as $T_{\rm MB}(3\mbox{--}2)-\xi T_{\rm MB}(1\mbox{--}0)$, where $\xi$ is the ratio between the total CO {\it J}=3--2 and CO {\it J}=1--0 intensities for quiescent gas ($\xi =0.57$).  We used the subtracted data as the CO {\it J}=3--2 wing, and thus the $\xi T_{\rm MB}(1\mbox{--}0)$ data became non-wing data.

\begin{figure}[htbp]
\epsscale{0.8}
\plotone{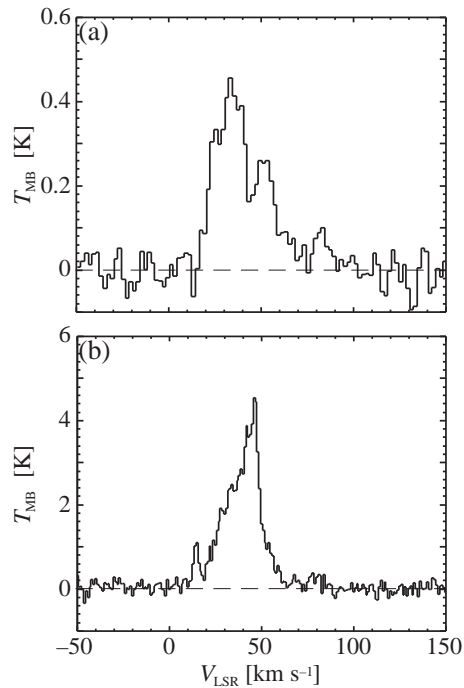}
\caption{Line profiles of ({\it a}) HCO$^+$ {\it J}=1--0 and ({\it b}) CO {\it J}=3--2 toward $(l, b)\!=\!(+34\dotdeg 82 , -0\dotdeg 47)$.  The velocity resolutions are 1.7 \kms\ and 0.43 \kms\ for HCO$^+$ and CO, respectively.  
\label{fig5}}
\end{figure}

\subsection{Expansion Model}
We employed a simple expansion model to describe the kinematics of the shocked gas.  The model consists of gas on the surface of a uniformly expanding prolate ellipsoid.  The center of the ellipsoid was taken at $(l, b)=(34\dotdeg 65, -0\dotdeg 39)$, and the position angle with respect to the Galactic plane was taken to be $70\arcdeg$.  The line-of-sight velocity can be written as 
\begin{eqnarray}
V_{\rm los} = \left\{ 
	\begin{array}{ll}
		V_{\rm sys} + V_{\rm exp}\, \sqrt{1 - s(x,y)^2}  & \left[ s(x,y) \leq 1 \right ] \\
		V_{\rm sys} &  \left[ s(x,y) > 1 \right ], \\
	\end{array} \right. 
\end{eqnarray}
where $V_{\rm sys}$ is the systemic velocity, and $V_{\rm exp}$ is the expansion velocity.  Further, $s(x, y)$ is the normalized projected distance from the center, defined as
\begin{equation}
1 - s(x,y)^2 \equiv \frac{1 - \left(\frac{x}{a}\right)^2 - \left(\frac{y}{b}\right)^2 }{1 + \frac{a^2-b^2}{a^2 b^2} x^2 } ,   
\end{equation}
where $a$ and $b$ are the semimajor and semiminor axes of the ellipsoid, and $x$ and $y$ are the projected distances from the center along the major and minor axes in the plane of the sky, respectively.  We employed $a=0\dotdeg 30$ and $b=0\dotdeg 21$ for the expanding shell of W44.  

The average velocity toward each position was calculated with each data sets using 
\begin{equation}
\left<V\right> = \int_{V_{\rm min}}^{V_{\rm max}} T_{\rm MB} V dV/ \int_{V_{\rm min}}^{V_{\rm max}}  T_{\rm MB} dV .
\end{equation}
The limits of integration were chosen to be $V_{\rm min}=+20$ \kms\ and $V_{\rm max}=+60$ \kms .  The systemic velocity was fixed at $V_{\rm sys}=47.5$ \kms , which is the average velocity for the HCO$^+$ and CO data at $s > 1$.  Figures 7 and 8{\it a} show $s(x, y)$--$\left<V\right>$ plots for HCO$^+$ and for the CO wing data.  A standard least-squares fitting procedure yielded $V_{\rm exp}=12.2\!\pm\!0.3$ \kms\ for HCO$^+$, and $V_{\rm exp}=13.2\!\pm\!0.2$ \kms\ for the CO wing data.  We employ an average value weighted by $1/\sigma^2$, $V_{\rm exp}=12.9\!\pm\!0.2$ \kms .  

\begin{figure}[htbp]
\epsscale{0.8}
\plotone{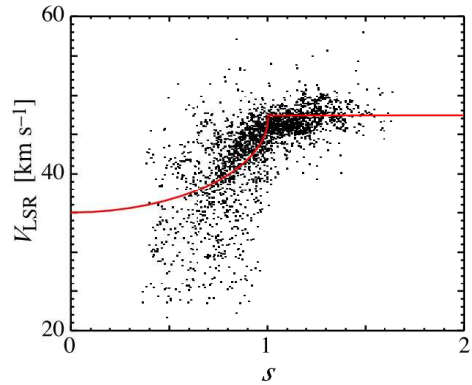}
\caption{Normalized distance versus average velocity [$s(x, y)$--$\left <V\right>$] calculated with the HCO$^+$ {\it J} = 1--0 wing data. 
\label{fig6}}
\end{figure}

\begin{figure}[htbp]
\epsscale{0.8}
\plotone{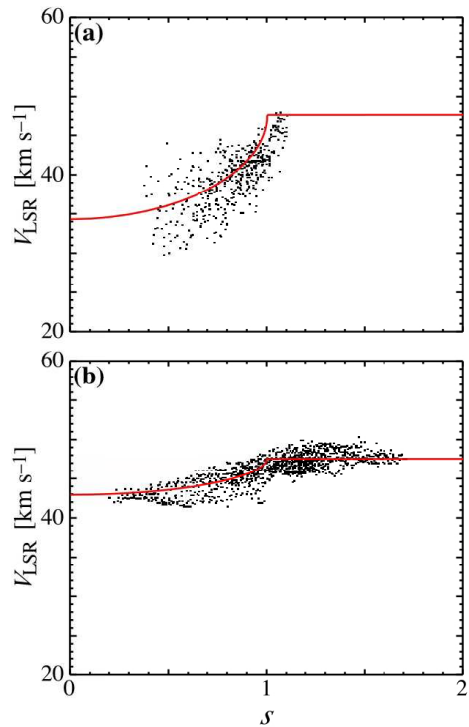}
\caption{({\it a}) Plot of normalized distance versus average velocity [$s(x, y)$-$\left <V\right>$] calculated with the CO {\it J}=3--2 ``wing" data.  ({\it b}) The $s(x, y)$-$\left <V\right>$ plot for the CO {\it J}=3--2 ``non-wing" data. 
\label{fig7}}
\end{figure}

\subsection{Kinetic Energy}
The mass of the shocked gas $M_{\rm s}$ is estimated from the HCO$^+$ {\it J}=1--0 intensity because it traces mostly shocked gas.  We assumed that the line is optically thin and the rotational levels are in local thermodynamic equilibrium (LTE) at $T_{\rm k}=40$ K.  We employed the abundance ratio $[{\rm HCO}^{+}]/[{\rm H}_{2}]\!=\!(2.9\!\pm\!1.5)\times 10^{-9}$, which is the average value for Galactic molecular clouds (Ungerechts et al. 1997; Liszt \&\ Lucas 2000).  We obtained $M_{\rm s}=(1.2\!\pm\!0.6)\times 10^{4}$ $M_{\sun}$, where the error includes the intensity calibration uncertainty ($\sim\!10 \%$).  Note that the estimated mass includes the SEMBE gas since we did not subtract the non-wing component from the HCO$^+$ data.  

Because of the finite optical depth [$\tau({\rm HCO}^{+})\sim 1$], $M_{\rm s}$ is revised upward by a factor of 2, whereas the subthermal conditions decreases it by a factor of several.  For instance, an LVG analysis with $n({\rm H}_2)=10^5$ cm$^{-3}$, $T_{\rm k}=40$ K, and $\tau({\rm HCO}^{+})\ll 1$ yields $M_{\rm s}=(3.6\!\pm\!1.9)\times 10^{3}$ $M_{\sun}$, which is 1/3 of the LTE estimation.  Since these effects cancels to some extent, we employed the LTE estimation in the following discussion.  The mass of shocked gas is a factor of 4 larger than that of SEMBE, $\sim 3000$ $M_{\sun}$, which was estimated from CO {\it J}=1--0 data assuming $T_{\rm k}=20$ K (Seta et al. 2004).  

The kinetic energy of the molecular gas provided by the SN blast wave consists of two terms, 
\begin{eqnarray}
\begin{array}{lll}
E_{\rm kin}({\rm mol}) & = & E_{\rm exp} + E_{\rm turb} \\
& = & \frac{1}{2} M_{\rm s} V_{\rm exp}^2 + \frac{3}{2} M_{\rm s} (\sigma_{\rm s}^2 - \sigma_{\rm q}^2) .   \\
\end{array}
\end{eqnarray}
The first term is the expansion energy, and the second corresponds to the increase in the turbulent energy, where $\sigma_{\rm s}$ is the one-dimensional velocity dispersion of the shocked gas, and $\sigma_{\rm q}$ is that of quiescent gas.  We obtained $\sigma_{\rm s}=6.8\!\pm\!0.1$ \kms\ and $\sigma_{\rm q}=2.30\!\pm\!0.02$ \kms\ using the wing and non-wing data for the CO {\it J}=3--2 line, respectively.  Thus, we obtained $E_{\rm exp}=(2.0\pm 1.1)\times 10^{49}$ erg and $E_{\rm turb}=(1.5\pm 0.8)\times 10^{49}$ erg, and thus $E_{\rm kin}({\rm mol})=(3.5\!\pm\!1.3)\times 10^{49}$ erg for the shocked molecular gas.  

In addition to the shocked molecular gas, an HI expanding shell is associated with the W44 SNR (Koo \&\ Heiles 1995).  Since this HI shell was detected in positive high velocities, it covers the far-side portion of the W44 SNR, while the GMC covers the near-side.  The lower limit to the total kinetic energy provided by the SNR is given by adding the kinetic energy of the HI shell, $(8.0\pm 1.0)\times 10^{49}$ erg, to $E_{\rm kin}({\rm tot})$, which yields $(1.2\!\pm\!0.2)\times 10^{50}$ erg.  On the other hand, the W44 GMC covers approximately $1/4$ of the radio shell in the plane of the sky.  Thus, the upper limit might be obtained by extrapolating $E_{\rm kin}({\rm mol})$ to the entire solid angle by multiplying by $8$, which yields $(2.8\!\pm\!1.1)\times 10^{50}$ erg.  


The kinetic energies obtained above correspond to a conversion efficiency of $\eta = (0.1\mbox{--}0.3) (E_{\rm bar}/10^{51}\, \mbox{erg})^{-1}$.  This is in good agreement with the prediction of an early theoretical study (Chevalier 1974) if the SN baryonic energy does not differ greatly from $10^{51}$ erg.  Note that the estimated upper and lower limits and the theoretical prediction are all compatible, despite uncertainties in our basic assumptions.

\subsection{Relation between Wing and SEMBE}
We analyzed the CO {\it J} = 3--2 non-wing data in the same way as the wing data.  Figure 7{\it b} shows an $s(x, y)$--$\left<V\right>$ plot for the CO {\it J} = 3--2 non-wing data.  A velocity gap at $s\simeq 1$ is apparent, and the behavior of $\left< V \right>$ is explained well by an expansion model.  A least-squares fitting yields $V_{\rm exp} = 4.7\!\pm\!0.1$ \kms\ for the non-wing gas.  This indicates that the non-wing gas is not totally quiescent, and that most of it is affected by the SNR.  This non-wing expanding gas might correspond to the SEMBE of the CO {\it J}=1--0 line (Seta et al. 2004).  

\begin{figure}[htbp]
\epsscale{1.0}
\plotone{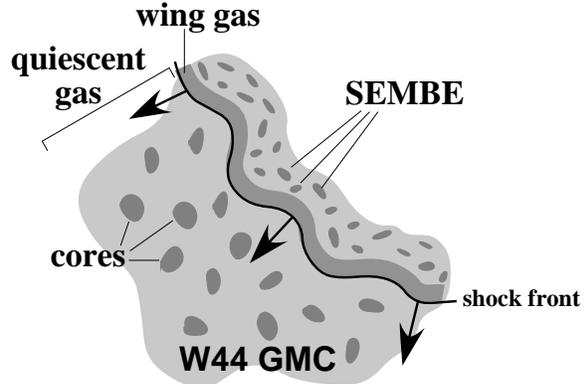}
\caption{Schematic view of W44 blast wave based on our HCO$^+$ {\it J} = 1--0 and CO {\it J} = 3--2 observations.  Non-wing gas is the mixture of quiescent and SEMBE gas.  
\label{fig8}}
\end{figure}

Although the nature of SEMBE is still controversial, its spatial distribution suggests that it is related to the blast wave of the W44 SNR.  The expansion of the non-wing gas might support this notion.  Its slow expansion and moderately broad velocity width indicate that it is located downstream of the wing-emitting gas with respect to the shock front.  Most likely it was a higher-density region before the passage of the SN blast wave because high-density gas decelerates shock waves.  A simple Sedov solution indicates that the expansion velocity is proportional to $(E/\rho)^{0.2}\,t^{-0.6}$ , where $t$ is the elapsed time from the explosion.  Thus, a preshock density that is higher by a factor of $240$ decreases the expansion velocity by a factor of $3$.   

Figure 8 shows a schematic view of the shocked gas adjacent to the W44 SNR.  We hypothesize that the precursor of the SEMBE gas consists of a number of unresolved dense clumps embedded in the W44 GMC.  These dense clumps decelerate SN blast wave distorting the shock front, and thereby being left behind it.  The SEMBE is barely detected in HCO$^+$ with $T_{\rm MB}\!\lesssim\!0.2$ K.  The complex configuration of the SEMBE gas, wing-emitting gas, and foreground quiescent gas generates the various observed shapes of the HCO$^+$ {\it J}=1--0 profiles.  

Size of putative dense clumps must be far smaller than $0.3$ pc in diameter, since they are unresolved by a $\sim 20\arcsec$ beam.  An LVG analysis with $n({\rm H}_2)\!=\!3\times 10^4$ cm$^{-3}$, $T_{\rm k}\!=\!20$ K, $[{\rm HCO}^{+}]/[{\rm H}_{2}]\!=\!2.9\times 10^{-9}$, and $T_{\rm MB}(\mbox{HCO}^{+})= 0.2$ K gives $N({\rm H}_2)/dV\!=\!1.6\times 10^{20}$ cm$^{-2}$ $(\kms)^{-1}$.  This column gives $T_{\rm MB}(\mbox{CO 1--0} )\simeq 8$ K, which is typical for the SEMBE, if we assume $[{\rm CO}]/[{\rm H}_{2}]\!=\!8\times 10^{-5}$.  Using $dV=5$ \kms , the size of dense clumps in the SEMBE is estimated to be $\sim 8\times 10^{-3}$ pc ($0.6\arcsec$ at 3 kpc).  ALMA may be able to resolve such a small clump at mm-wavelength.  

Moderately broad ($\Delta V=7.5$ \kms ) CO emission was detected also from the SNR G18.8+0.3 (Dubner et al.  2004).  This could be another example of the SEMBE.  More SEMBE detections would be made by careful inspection of the existent CO data and by sensitive molecular line observations toward SNR-MC interacting systems (e,g., Table 2 in Jiang et al. 2010).  Direct imagings of the dense clumps reponsible for SEMBE using millimeter  and submillimeter arrays would be essential to examine the SEMBE model.

\section{Conclusion}
Observations by the NRO 45 m and ASTE 10 m telescopes toward the GMC associated with W44 yielded the following conclusions.  
\begin{enumerate}
\item{Extensive HCO$^+$ {\it J} = 1--0 and CO {\it J} = 3--2 maps delineated the distribution and kinematics of shocked molecular gas adjacent to the W44 SNR.}
\item{The average velocity distributions of the shocked gas can be fitted by a uniform expansion model.  The best-fit expansion velocities are $12.2\!\pm\!0.3$ \kms\ and  $13.2\!\pm\!0.2$ \kms\ in HCO$^+$ and CO, respectively.}
\item{The non-wing CO {\it J}=3--2 component is also fitted by the same model with an expansion velocity of $4.7\!\pm\!0.1$ \kms .  This component might be dominated by SEMBE.  The precursor of the SEMBE gas might be a number of unresolved dense clumps that decelerated the SN blast wave.}
\item{The estimated mass and kinetic energy of the shocked molecular gas were $M_{\rm s}=(1.2\!\pm\!0.6)\times 10^{4}$ $M_{\sun}$ and $E_{\rm kin}({\rm mol})=(3.5\!\pm\!1.3)\times 10^{49}$ erg, respectively.}
\item{After adding the energy of the HI shell, we concluded that at least $(1.2\!\pm\!0.2)\times 10^{50}$ erg has been converted into gas kinetic energy from the initial baryonic energy of the W44 SN.  On the other hand, the upper limit to the energy available for disturbing interstellar matter was $(2.8\!\pm\!1.1)\times 10^{50}$ erg.}
\item{We also found ultra-high-velocity CO {\it J}=3--2 wing emission with a velocity width of $\sim 100$ \kms\ at $(l, b)\!=\!(+34\dotdeg 73, -0\dotdeg 47)$.  The origin of this extremely high-velocity wing is quite mysterious.}
\end{enumerate}

\acknowledgments
We are grateful to the NRO staff for their excellent support of the 45 m observations.  The Nobeyama Radio Observatory is a branch of the National Astronomical Observatory of Japan, National Institutes of Natural Sciences.  We thank the members of the ASTE team for the operation of the telescope and ceaseless efforts to improve the ASTE. Observations with ASTE were conducted remotely from the NRO using NTT's GEMnet2 and its partner R\&E (Research and Education) networks, which are based on the AccessNova collaboration between the University of Chile, NTT Laboratories, and the National Astronomical Observatory of Japan.  We also thank the anonymous referee for helpful comments and suggestions to improve the manuscript.  This work is partially supported by JSPS Grant-in-Aid for Scientific Research(C) No. 24540236.    

\end{document}